# V2X communication coverage analysis for connected vehicles in intelligent transportation networks: A case study for the city of Xanthi, Greece

E. Bazinas, A. Gregoriades, M. Raspopoulos, M. Georgiades

*Abstract* — Intelligent transportation systems (ITS) have been developed to improve traffic flow, efficiency, and safety in transportation. Technological advancements in communication such as the Vehicle-to-Everything (V2X), Vehicle-to-Vehicle (V2V) and Vehicle-to Infrastructure (V2I) enable the real-time exchange of information between vehicles and other entities on the road network, and thus play a significant role in their safety and efficiency. This paper presents a simulation study that models V2V and V2I communication to identify the most suitable range of data transmission between vehicles and infrastructure. The provincial city of Xanthi, Greece is used as a cases study, and the goal is to evaluate whether the proposed placement of Road Side Unit (RSU) provided adequate communication coverage on the city's road network. An analysis through different scenarios identified improvements in traffic management, driving behavior and environmental conditions under different RSU coverage. The results highlight that the communication range of 400 meters is the most adequate option for optimum traffic management in the city of Xanthi.

*Index Terms*— Smart City, Intelligent Transportation Systems, Connected Vehicles, VEINS, OMNeT++, SUMO, V2V, V2I, VANET, RSU.

## I. INTRODUCTION

In the current era of rapid technological advancement, the concept of "Smart City" has become a major focus of interest for many countries globally and has become a popular topic of research among scholars. Among the domains of Smart City, Intelligent Transportation Systems (ITS) has garnered significant attention. An emerging technology within the realm of ITS is Vehicle-to-Everything (V2X) communication, which facilitates communication between vehicles and other entities in their environment, including infrastructure, other vehicles, pedestrians, and devices. Vehicle-to-Vehicle (V2V) and Vehicle-to-Infrastructure (V2I) communication, which are integral components of the V2X system, enable the exchange of data and warning messages, allowing drivers to receive information on potential situations that may impact traffic on the road network, such as accidents.

As mentioned in [1] and [2], V2V and V2I communication technology has been recognized as a vital component in intelligent transportation systems, facilitating wireless communication between vehicles and infrastructure to enable the exchange of warning messages and information packets. These serve to inform drivers about potential issues on the road network such as accidents, weather conditions and traffic congestion, providing them with real-time information to select alternate routes to reach their destination expeditiously. This specific technology is not applied to a fixed structure, but is instead based on wireless communication between vehicles and RSUs to ensure the functionality of the network [3].

Road Side Units (RSUs) are an essential component of vehicular communication systems that provide connectivity to vehicles traveling on roads [4]. They are typically installed along the roadsides, at intersections, and other locations where vehicles need to communicate with each other and with the infrastructure. RSUs typically consist of a radio transceiver, a processing unit, and a power source. They use wireless communication technologies such as IEEE 802.11p (also known as WAVE) to communicate with nearby vehicles and other RSUs. An RSU can provide a wide range of services to vehicles, such as traffic information, collision warnings, and emergency response services.

In the context of a Smart City, numerous technologies and applications have been developed to enhance the quality of life of citizens. One significant area that is positively impacted is traffic management on the city's road network, through the implementation of intelligent systems, aimed at preventing accidents and reducing congestion phenomena especially in emergency situations. Unfortunately, there have been several instances where emergency vehicles, such as ambulances, fire brigades and Police vehicles have been unable to reach the scene of the incident on time due to traffic problems, resulting in precious time being lost. Thus, the need for this study in the Greek provincial city of Xanthi cannot be overstated, given the unique features of its road network coupled with the driving behavior of its inhabitants, which often have a detrimental impact on traffic management, especially in emergency situations.

To facilitate the implementation of the intelligent transportation network with connected vehicles, in this study we used sophisticated software tools such as VEINS (Vehicular Network Simulation) [4], SUMO (Simulation of Urban Mobility) [5] and OMNET++ (Objective Modular Network Testbed in C++) [9]. A combination of these software tools has been used by many researchers worldwide to create and perform simulations based on realistic scenarios. These powerful tools provided a robust platform to simulate various aspects of communication and vehicle mobility patterns, allowing the study to derive valuable insights into the efficiency and effectiveness of V2X communication systems in enhancing traffic management in the urban environment of the city of Xanthi.

## II. RELATED WORK

### A. Simulation-Based Case Studies

Authors in [5] developed a highly realistic scenario, called the Luxembourg SUMO Traffic (LuST) scenario, which is a realistic representation of mobility patterns in the city of Luxembourg and its surroundings. It includes a 24-hour period of vehicle traffic, with a focus on the morning and afternoon peak hours. They also provided an evaluation of the scenario using a number of vehicular networking applications, including vehicle-to-vehicle (V2V) and vehicle-to-infrastructure (V2I) communication, as well as mobility prediction and routing.

Paper [6] focused on the simulation of Cooperative Intelligent Transport Systems (C-ITS) for multimodal transportation in the city of Monaco using SUMO. The paper discussed the challenges and requirements for developing a simulation framework for C-ITS that integrated multiple modes of transportation, including pedestrians, bicycles, cars, and buses. The authors described the traffic scenario in Monaco and how they modeled it using SUMO, including the infrastructure, traffic regulations, and travel demand.

Authors in [7] proposed a decentralized approach for vehicle re-routing using vehicular ad-hoc networks (VANETs) in the case of unexpected events such as road closures or accidents. The proposed approach aimed to reduce the overall travel time and improve the traffic flow by re-routing the vehicles in a decentralized manner using information provided by other vehicles in the network. The paper first introduced the basic concepts of VANETs and the proposed decentralized approach for vehicle rerouting. The authors then presented the simulation setup using parts of Colombo and Kandy cities in Sri Lanka as study areas for their experiment. Finally, they evaluated the performance of the proposed approach using SUMO, OMNeT++ and VEINS. The simulation results demonstrated that the proposed approach could reduce the average travel time and increase the overall network throughput, even under high traffic densities and network loads.

Paper [8] presented a study on the development of a realistic urban road traffic simulation model for the city of Erbil, Iraq. Authors used software such as VEINS, SUMO, OMNet++, PTV Vissim and PTV Visum to develop a simulation model with bidirectional coupling. The study first involved collecting and analyzing traffic data for Erbil City, including vehicle counts, travel times, and road network geometry. The authors then calibrated and validated the simulation model using the collected data to ensure that it accurately represented the traffic conditions in the city. The simulation model was then used to analyze the impact of various traffic management strategies on traffic flow and congestion in the city. The authors evaluated the effectiveness of strategies such as traffic signal optimization, bus lane implementation, and intelligent transport systems. The results showed that the simulation model was able to accurately represent the traffic conditions in Erbil City, and that the proposed traffic management strategies had a significant impact on reducing traffic congestion and improving traffic flow.

A practical application of the OMNeT++ simulator, which is an integrated graphical environment for developing and running simulations by accessing libraries and frameworks developed in the C++ programming language to create network simulations, was presented in [9].

In [10] the authors, who emphasized the necessity of using an efficient routing protocol that meets the requirements and characteristics of VANETs in order to achieve good performance in terms of average packet loss rate and average end-to-end packet delay, used the same software combination to test a new routing protocol (ProMRP) specifically designed for VANETs. The proposed protocol aimed to improve the efficiency and reliability of routing by considering multiple metrics such as delay, packet loss, and signal strength. The protocol also took into account the uncertainty of VANETs by using probability-based decision-making to select the most suitable next hop. The authors evaluated the performance of the proposed protocol using a realistic urban mobility model and compared it with three other popular VANET routing protocols. The simulation results showed that the proposed protocol achieved better performance in terms of packet delivery ratio, end-to-end delay, and network throughput, especially in high-density urban scenarios.

### B. RSU Deployment Investigations

An investigation of the impact of real-time path planning on reducing vehicle travel time in urban traffic networks was carried out by the authors in [11], who proposed a real-time path planning algorithm that took into account real-time traffic conditions and estimated the travel time for each possible path. The algorithm used data from RSUs and other sources to collect and transmit traffic data such as vehicle speed, density, and congestion levels in order to compute the optimal path for each vehicle to reach its destination in the shortest time possible while avoiding congested areas. Performance evaluation of the proposed algorithm was carried out through simulation experiments using SUMO software. The simulation results showed that the proposed algorithm could significantly reduce vehicle travel time compared to traditional shortest path algorithms and had significant implications for reducing congestion, improving mobility and enhancing the sustainability of urban transport systems.

The successful implementation of V2V and V2I communication is dependent upon the installation of RSUs at multiple locations throughout the road network, as well as the availability of On-Board Units (OBUs) in vehicles, as discussed in [12]. In the same research, the authors proposed a novel approach to optimize the placement of RSUs in V2I networks using a genetic-based algorithm that took into account realistic factors such as road network topology, traffic flow, and communication range. The proposed algorithm aimed to maximize network coverage and minimize communication delay by strategically placing RSUs in the network. The authors conducted extensive simulations to evaluate the effectiveness of their proposed approach, comparing it to other existing methods. The simulation results showed that the proposed approach outperformed other methods in terms of coverage, delay, and cost.

Paper [4] provided a comprehensive survey of the literature on RSU deployment in Internet of Vehicles (IoV) systems. The authors first introduced the concept of IoV and the role of RSUs in IoV systems. They then reviewed the different types of RSUs and their communication capabilities, as well as the various deployment strategies for RSUs. The authors also discussed the challenges and open issues related to RSU deployment in IoV systems, such as the optimal placement of RSUs, the selection of RSU communication technologies, and the management of RSU resources.

Authors in [13] proposed an optimized RSU deployment mechanism in VANETs that was designed to prevent data loss. The proposed mechanism used an efficient protocol to manage data transmission between vehicles and RSUs, which reduced the risk of data loss and improved the overall reliability of the network. The authors performed detailed simulations to evaluate the performance of the proposed mechanism by comparing it with other existing methods. The simulation results showed that the proposed mechanism outperformed other methods in terms of data loss prevention, network reliability, and communication overhead.

In [14] authors proposed a method for deploying RSUs in V2X networks to improve traffic prediction accuracy. They used a machine learning approach to predict traffic flow using real-time data from RSUs and vehicles, and then used optimization techniques to determine the optimal locations for the RSUs. The paper evaluated the proposed approach using real-world traffic data from the city of Budapest and compared it with other existing approaches. The results showed that the proposed method outperformed existing approaches in terms of prediction accuracy and RSU deployment efficiency. The paper concluded that the proposed method could be useful in improving traffic management and reducing congestion in urban areas.

Paper [15] investigated the effect of RSU range on the performance of vehicle-to-vehicle (V2V) and vehicle-to-infrastructure (V2I) communication in VANETs. The author used simulation to evaluate the impact of different RSU ranges on communication reliability, transmission delay, and network throughput. The simulation results showed that the communication reliability and network throughput increased as the range of RSUs increased. However, the transmission delay also increased with larger RSU ranges, indicating a trade-off between reliability and delay. The findings could inform the development of more effective RSU deployment strategies that balance communication reliability and delay.

An RSU placement scheme for VANETs aimed at improving network connectivity and coverage is proposed by the authors in [16]. The proposed scheme used a combination of heuristic and optimization algorithms to identify the optimal locations for RSUs based on traffic density and network coverage requirements. Simulation results showed that the proposed system outperformed other existing methods in terms of network connectivity, coverage, and communication delay and could have a significant impact on improving traffic safety and reducing congestion by improving the efficiency of communication between vehicles and between vehicles and infrastructure.

An RSU-assisted emergency vehicle transit approach for urban roads using VANETs was presented in [17]. The authors' approach used RSUs to identify emergency vehicles and prioritize their transit through urban roads by giving them green light priority. The authors evaluated the performance of their proposed approach through simulation experiments using the Network Simulator-2 (NS-2) platform. The simulation results showed that the proposed approach could significantly reduce emergency vehicle transit time and improve traffic flow efficiency.

The authors in [18] proposed an efficient V2X-based vehicle tracking method that used a single RSU and a single receiver. The proposed method used the time difference of arrival (TDOA) and received signal strength (RSS) of the wireless signals received by the RSU and the receiver to estimate the vehicle location. They evaluated the performance of the proposed method through simulation experiments using MATLAB software. The simulation results showed that the proposed method achieved high localization accuracy with low computational complexity and communication overhead.

A new routing protocol for urban VANETs, which incorporates the use of RSUs to enhance communication and traffic awareness, was implemented in [19]. The proposed protocol used a reinforcement learning (RL) approach to optimize routing decisions by considering the current traffic state and the presence of RSUs. The RL algorithm learned from historical data to predict the best route to reach a destination, considering the traffic conditions and the locations of RSUs. The authors evaluated the proposed protocol using simulation experiments and the results showed that it could achieve better performance in terms of packet delivery ratio, end-to-end delay, and routing overhead compared to traditional routing protocols. The paper provided a detailed explanation of the proposed protocol, the RL algorithm used, as well as the simulation setup and results.

Authors in [20] proposed an RSU deployment strategy for urban vehicular networks with limited cost and delay. The authors emphasized the importance of RSUs as a key communication infrastructure for V2V and V2I communication in VANETs. They presented a mathematical model for determining the optimal RSU deployment strategy that minimized the total deployment cost while meeting the delay constraints. The model considered parameters such as vehicle density, RSU transmission range, and communication delay requirements. The proposed deployment strategy was evaluated through simulation experiments, and the results showed that it outperformed other existing strategies in terms of both delay performance and deployment cost.

Paper [21] presented a new approach for vehicular communication in urban environments, proposing a reinforcement learning-based routing algorithm for VANETs. The algorithm was designed to improve the efficiency of V2X communication by selecting the most appropriate routing paths based on intersections in urban areas. The proposed approach used a deep Q-network (DQN) to learn the optimal intersection-based routing policies for V2X communication. The DQN was trained on a dataset of simulated vehicle traffic scenarios and the resulting policies were evaluated in terms of communication efficiency, packet delivery ratio, and end-to-end delay. The paper provided experimental results that demonstrated the effectiveness of the proposed algorithm compared to other state-of-the-art routing algorithms in V2X communication. The proposed algorithm performed significantly well over other approaches in terms of packet delivery ratio, communication efficiency, and end-to-end delay.

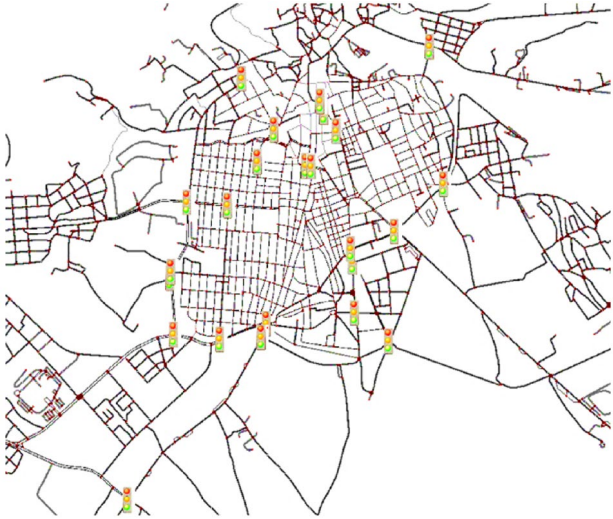

Figure 1. Locations of traffic lights on the city map based on information received from Municipality and Traffic Police Department of Xanthi.

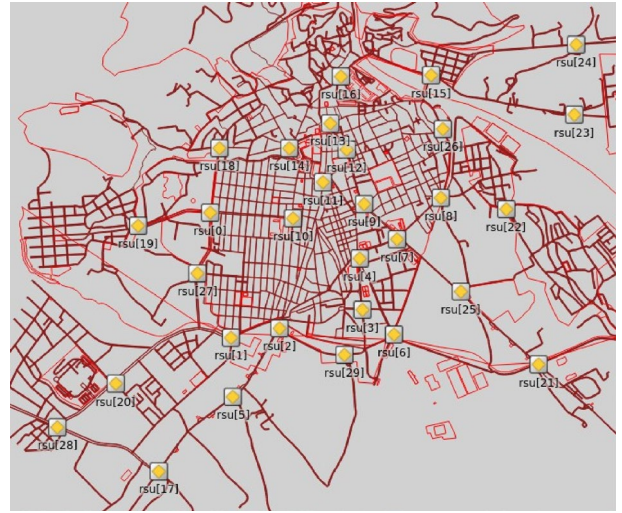

Figure 2. Proposed locations of RSU units on the city map of Xanthi considered for simulation.

## III. SIMULATION CONFIGURATION

As mentioned in [22], VEINS is an open-source simulation framework for vehicular networks. The framework is built on top of the widely-used network simulator OMNeT++ and the road traffic simulator SUMO. It allows for the simulation of realistic vehicular communication scenarios in urban and highway environments, including different types of vehicles and communication technologies. An extensive description of the VEINS architecture, including the integration of OMNeT++ and SUMO, as well as the implementation of various communication protocols and applications for vehicular networks, is provided by the authors.

In this research paper we used that framework to evaluate various RSU coverage configurations. To achieve this, we conducted a series of simulations involving 1,220 vehicles appearing at random intervals and following specific routes within the road network of the city of Xanthi and 30 RSUs placed at various locations on the map. The simulations were performed for a duration of 4,600 seconds. We programmed 47 of the vehicles to randomly stop due to events, such as accidents, each of which lasted 300 seconds. During this immobilization period, these vehicles continuously transmitted necessary warning messages to nearby RSUs and vehicles. In order to establish communication between vehicles and infrastructure in the simulations conducted, the IEEE 802.11p communication standard was employed. This standard was specifically designed to facilitate high-speed and low-latency communication between vehicles and RSUs [23], and is an integral component of the VEINS framework. In addition, we utilized SUMO and Netedit software tools to make appropriate adjustments to the smart traffic lights. Specifically, we converted the phase change from static to dynamic, such that the traffic lights adjusted the duration between phases based on the traffic congestion in the road network, as detected by sensors. The simulation parameters used in this study are detailed in Table I.

In order to ensure that communication coverage of the entire road network throughout the city of Xanthi was sufficient, strategic placements of RSU units were carefully considered. Specifically, the decision was made to position these units at locations where traffic lights are present on the road map, as visually demonstrated in Figures 1 and 2.

TABLE I. Parameters used in the Simulations

| Parameters | Values |
|---|---|
| Simulation Region | City of Xanthi |
| Total Area | 13,60 km² |
| Number of Vehicles | 1.220 |
| Number of RSUs | 30 |
| Communication Range | 100 – 600 meters |
| Simulation Duration | 4600 seconds |
| Event Start Time | Values between the 153rd and 350th second after the vehicle entry into the simulation |
| Event Duration | 300 seconds |
| Communication Standard | IEEE 802.11p |
| Antenna Type | SampledAntenna1D |
| Transmission Power | 20 mW |
| Transmission Rate | 6 Mbps |
| Min Power Level | -110 dBm |
| Noise Floor | -98 dBm |

Our decision was based on extensive data obtained from various sources. The Municipality and traffic police department of Xanthi provided data concerning the frequency of traffic congestion, traffic accidents and the corresponding issues addressed in traffic management at places where traffic lights were present. This information was used to select the loca-tions for RSU placement. Additionally, certain locations exhibiting a high density of vehicles and frequent instances of traffic congestion were also deemed suitable for RSU unit deployment. Through these selections, we aimed to maximize the communication range of the RSUs and ensure the effectiveness of the vehicular network in improving traffic management and overall driving behavior.

## IV. SIMULATION RESULTS

### A. Data Packet Loss

Based on the results obtained from the simulations, we observed that the number of warning messages exchanged during scheduled accidents, between vehicles and RSUs, increased as the communication range expanded due to the increased number of vehicles and RSUs within range at the time of transmission, as shown in Figure 3a and Figure 3b.

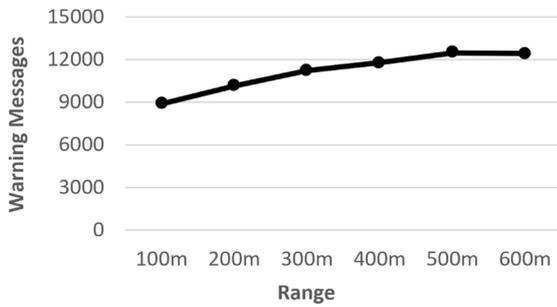

Figure 3a. Total number of warning messages received.

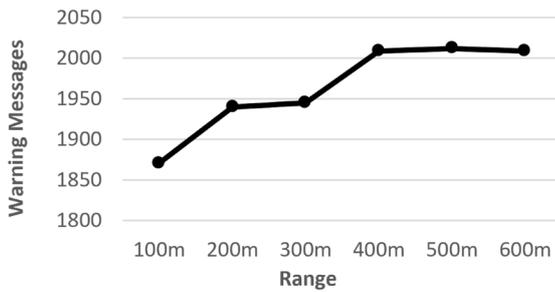

Figure 3b. Total number of warning messages sent.

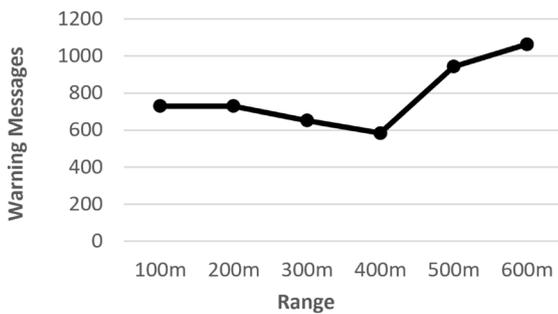

Figure 3c. Total number of warning messages lost.

However, the increase in range had a negative impact on data loss by significantly increasing the number of lost warning messages during simulations with communication range between 500 and 600 meters as shown in Figure 3c.

This is due to the fact that as the communication range between vehicles and RSUs increased, the obstacles and interference in the signal path also increased. Given that the power control was disabled during the simulations, the increase in range resulted in attenuation, reflection and refraction of the signal, leading to a further increase in the loss of warning messages. In addition, as the communication range increased, the signal strength decreased and noise and interference from other sources became more significant. This affected the quality of the signal and increased the possibility of data loss. In addition, with the increase in communication range, there were more nodes or vehicles to be served, which led to greater network congestion. This contributed to further increase in packet loss which in turn caused undesired delays and negative overall performance.

*B. Test Vehicles Results*

In order to evaluate potential improvements in traffic management and driving behavior of the vehicles, we integrated six test vehicles into the simulations. These vehicles followed predefined routes on the city road map, in terms of departure and arrival points, during the simulations. Our aim was to record the changes in travel time of each test vehicle and to evaluate the possible improvements resulting from the readjustment of vehicles' route to the destination after receiving warning messages. Total travel time for each test vehicle to get from the departure point to the arrival point, during simulations with different range coverages, is shown in Figure 4. Through our analysis, it was discovered that across all test vehicles, the simulation with a communication range of 400 meters yielded the lowest vehicle travel time. The adoption of this particular range within the simulation resulted in expedited arrival at the destination by the test vehicles, effectively bypassing traffic congestion arising from incidents.

Based on the results gathered from the six test vehicles, we observed that V2V and V2I communication technology facilitated the exchange of critical information about traffic conditions, accidents and potential hazards on the city's road network. This allowed vehicles to make informed decisions about their driving behavior and route choice. The information received in real-time via V2V and V2I communication helped vehicles to avoid congested routes and choose alternative routes, thus reducing travel time.

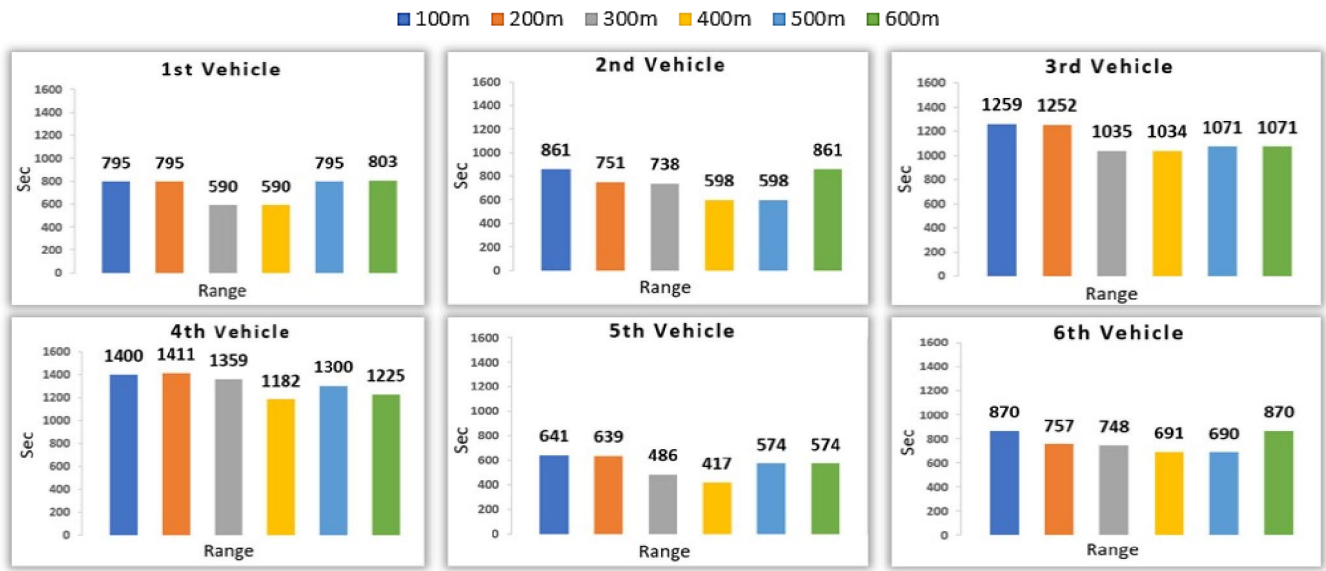

Figure 4. Total travel time for each test vehicle under different range coverages.

TABLE II. Average values recorded by all vehicles during simulations of different scenarios in terms of communication capabilities, accident frequency and range coverages.

| | Best Case Scenario | 100m | 200m | 300m | 400m | 500m | 600m | Worst Case Scenario |
|---|---|---|---|---|---|---|---|---|
| Total Time (seconds) | 514 | 773 | 640 | 568 | 551 | 670 | 720 | 833 |
| Total Distance (meters) | 7251 | 7323 | 7431 | 7591 | 7630 | 7520 | 7380 | 7251 |
| Total speed (miles/hour) | 14,1 | 9,5 | 11,6 | 13,4 | 13,8 | 11,2 | 10,3 | 8,7 |
| Total $CO_2$ (micrograms) | 2,055 | 3,094 | 2,560 | 2,272 | 2,204 | 2,680 | 2,880 | 3,332 |

The results revealed that the transmission range during V2V and V2I communication plays a crucial role in transmitting warning messages and providing accurate information to vehicles. The simulation results showed that when a range of 400 meters was applied, vehicles received more accurate information, allowing them to adjust their route to the destination in a timely and efficient manner.

## V. OVERALL IMPROVEMENTS

In order to evaluate the effectiveness of V2V and V2I communication technology in the road network of the city of Xanthi, we conducted a comparative analysis to measure the improvement of total vehicle driving time, total distance travelled, total maintained speed and total $CO_2$ levels emitted by the vehicles. To achieve this, two new simulations were conducted to reproduce actual traffic conditions existing at the moment in the city of Xanthi. The first simulation, which served as a reference scenario, included the worst-case scenario where all intelligent systems implemented in previous simulations were disabled. In this simulation there was no communication between vehicles and RSUs, vehicle course change functions were disabled and static phase changes of traffic lights were used. In addition, the same number of vehicles and scheduled events, causing traffic problems on the road network, were maintained.

For the second simulation, the best-case scenario was applied, which maintained the absence of intelligent systems, except that there were no events on the road network that could affect vehicle traffic and arrival time at the destination. The overall results are illustrated in Table II. The analysis of these results allowed us to evaluate the overall effectiveness of V2V and V2I communication technology in mitigating traffic problems and improving travel time at a local perspective.

### A. Total Travel Time

In relation to the overall travel time of vehicles, the results show that the simulation with a communication range of 400 meters achieved the lowest overall vehicle travel time, with an average value of 551 seconds as shown in Figure 5a. This finding is significant as it shows a 33.9% reduction in travel time compared to the worst-case scenario values as shown in Figure 5b.

It is evident that the V2V and V2I communication technology used in the simulations facilitated the exchange of critical information about traffic conditions, accidents, allowing vehicles to make informed decisions about their routes and driving behavior. As a result, the simulation with a range of 400 meters allowed vehicles to avoid congestion and maintain a steady traffic flow, reducing travel time.

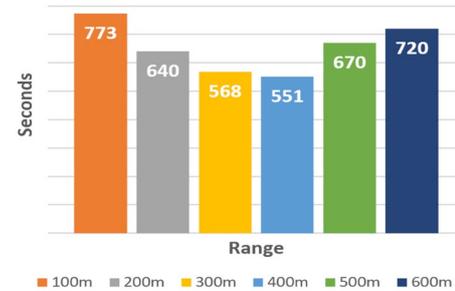

Figure 5a. Average total travel time of all vehicles under different range coverages.

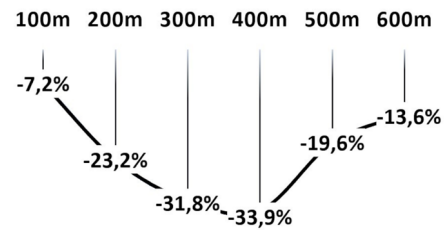

Figure 5b. Travel time reduction in relation to worst-case scenario based on Figure 5a.

### B. Total Distance

The simulation results showed that the total distance travelled by the vehicles was greater when V2V and V2I communication technology with a transmission range of 400 meters was applied as shown in Figure 6a. In particular, the average value of this parameter was found to be 7.630 meters. This increase in distance travelled, which was about 5% compared to the worst-case scenario values as shown in Figure 6b, is attributed to the ability of vehicles to change route in order to avoid traffic congested areas. The implementation of the V2V and V2I communication system facilitated the exchange of relevant information on traffic conditions and accidents, allowing vehicles to make informed decisions on the optimal route to their destination. The increase in distance travelled by vehicles is justified by the readjustment of their route to the final destination to avoid congestion and possible delays caused by accidents. As a result, the vehicles covered a longer distance to the final destination due to the readjustment of the route yet reached their destination with less delay compared to the delay in the worst-case scenario.

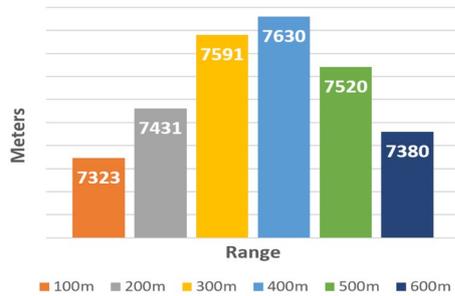

Figure 6a. Average total distance covered by all vehicles under different range coverages.

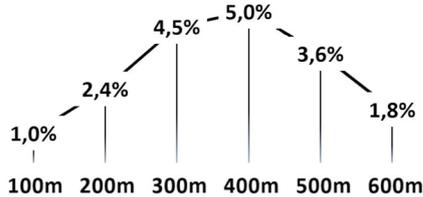

Figure 6b. Distance increase in relation to worst-case scenario based on Figure 6a.

### C. Total Vehicle Speed

Vehicle speed in the simulation with a communication range of 400 meters showed an overall increase, with an average value of 13.8 m/h as shown in Figure 7a. This result indicates a 36.8% increase in speed compared to the speed values of the worst-case scenario, as depicted in Figure 7b, indicating that congestion was avoided and fewer vehicles were brought to a standstill. The implementation of V2V and V2I communication enabled the timely exchange of relevant traffic information, allowing vehicles to make informed decisions and choose less congested routes, thus increasing overall vehicle speed.

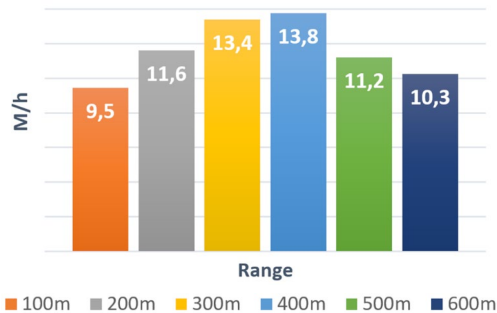

Figure 7a. Average total speed maintained by all vehicles under different range coverages.

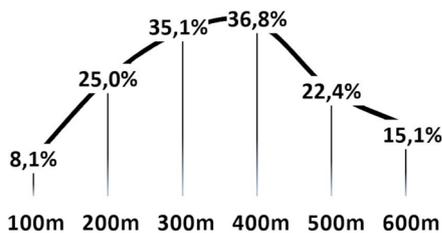

Figure 7b. Speed increase in relation to worst-case scenario based on Figure 7a.

### D. Total CO2 Emission Levels

In the simulation with a communication range of 400 meters, the lowest $CO_2$ emission levels from the vehicles were observed, with an average emission value of 2,204 micrograms as shown in Figure 8a. Comparison with the values of the worst-case scenario reveals a reduction of 33.9%, which is depicted in Figure 8b, confirming the effectiveness of the communication technology in reducing $CO_2$ emissions.

The reduction in emissions can be attributed to the re-routing of vehicles to their destination, thus avoiding congestion caused by accidents and preventing stop-and-go driving, which leads to higher $CO_2$ emissions. Timely rerouting of vehicles results in smoother driving conditions, which leads to reduced $CO_2$ emissions. These findings underscore the significance of V2V and V2I communication in realizing the objectives of intelligent transportation systems, while concurrently yielding favorable environmental outcomes.

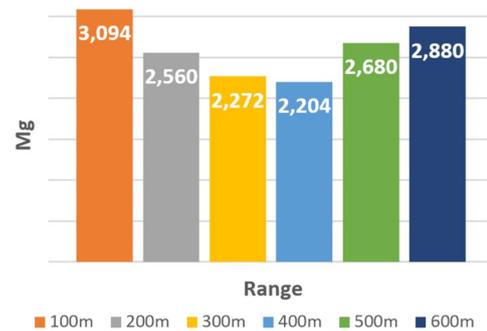

Figure 8a. Average total $CO_2$ level emitted by all vehicles under different range coverages.

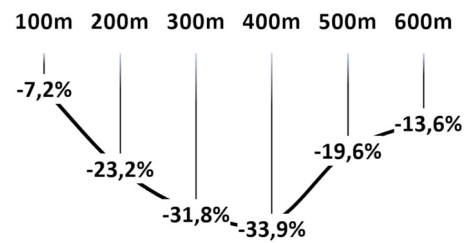

Figure 8b. $CO_2$ emission decrease in relation to worst-case scenario based on Figure 8a.

## VI. CONCLUSION

After conducting a thorough analysis of the simulation results, we observed that the implementation of V2V and V2I communication systems in the road network of the city of Xanthi led to a significant improvement. The structural design and architecture of the city, together with the condition of the road network, have been taken into account when evaluating the effectiveness of the communication systems.

For this paper, a series of simulations were performed to assess the efficacy of different communication ranges between vehicles and RSUs, with values ranging from 100 to 600 meters. Our findings indicate that the communication range of 400 meters is the most effective in providing adequate coverage for the entire road network. The selection of RSUs placement locations played a critical role in determining the optimal communication range. Communication ranges of 100 to 300 meters were found to be inadequate, based on the positioning of the RSUs on the city map, while ranges of 500 and 600 meters led to significant signal overlap and hindered the successful exchange of warning messages. Furthermore, the simulation results showed that the loss rate of warning messages was significantly reduced by applying the 400-meter communication range, indicating the provision of valid information for vehicles. This outcome contributed to increased options for vehicles to adjust their initial route to their destination.

Moreover, the adoption of the communication range of 400 meters resulted in a decrease in the occurrence of traffic congestion associated with accidents, as compared to the levels of congestion observed at other ranges. This reduction led to improved traffic management on the road network.

Additionally, a decrease in the travel time required for vehicles to reach their destination was observed when using a communication range of 400 meters, as compared to other ranges. This finding can be attributed to the more efficient coverage of the road network achieved at this range, which allowed for successful message exchange and, consequently, provided vehicles with more route adjustment options.

Furthermore, it was observed that the implementation of a 400-meter communication range resulted in a reduction of $CO_2$ emission levels compared to other ranges emitted by vehicles, thus indicating a favorable environmental impact. This reduction in $CO_2$ emissions can be attributed to the prompt adjustment of vehicle routes in response to the transmitted product messages. This readjustment facilitated the avoidance of traffic congestion caused by accidents, thereby preventing the vehicles from coming to a standstill and leading to a reduction in the release of $CO_2$ emissions into the atmosphere.

Based on the findings clarified above, as well as the recognition of the advantages that can result from the support of V2V and V2I communication systems, we believe that the adoption of intelligent transportation systems in the urban centre of Xanthi is imperative. Such a measure has the potential to significantly improve the quality of life of the city's residents and, therefore, deserves serious consideration by the relevant authorities.

## VII. FUTURE WORK

Potential future work could focus on the integration of reinforcement learning-based intelligent routing algorithms into the intelligent transport system to optimize traffic flow and alleviate congestion. These algorithms could leverage traffic prediction models to identify potential congested areas and proffer alternative routes to drivers, thereby circumventing congested roads and reducing travel time. Within the ambit of intelligent transportation systems, reinforcement learning could be utilized to streamline traffic flow and minimize congestion by dynamically adjusting traffic signal timing based on real-time traffic data, weather conditions, and other pertinent factors.

Furthermore, emergency vehicles, such as ambulances, fire trucks, and patrol cars, could be integrated into the V2X communication network, enabling them to communicate their location and status with other vehicles and infrastructure. This integration could permit emergency vehicles to receive priority access to intersections, alter traffic signal patterns, and communicate their route to other vehicles in real-time, ensuring their prompt and secure arrival at their destination. Additionally, reinforcement learning could be harnessed to optimize emergency vehicle routing by drawing insights from historical data to make more accurate predictions regarding the fastest and safest routes for emergency vehicles.

Overall, the integration of Reinforcement Learning-based intelligent routing algorithms, traffic prediction, V2X communication and emergency vehicle management can further improve the efficiency and effectiveness of the intelligent transport system, leading to reduced congestion, improved traffic flow and a better quality of life for the residents of a Greek city regardless of scale. Finally, the authors plan to undertake similar investigations in other cities as well with real traffic data for comparison purposes.

REFERENCES

[1] I. A. Aljabry and G. A. Al-Suhail, "A Survey on Network Simulators for Vehicular Ad-hoc Networks (VANETS)," *Int. J. Comput. Appl.*, vol. 174, no. 11, pp. 1–9, 2021, doi: 10.5120/ijca2021920979.

[2] M. Lee and T. Atkison, "VANET applications: Past, present, and future," *Veh. Commun.*, vol. 28, p. 100310, Apr. 2021, doi: 10.1016/j.vehcom.2020.100310.

[3] M. M. Hamdi, L. Audah, S. A. Rashid, A. H. Mohammed, S. Alani, and A. S. Mustafa, "A Review of Applications, Characteristics and Challenges in Vehicular Ad Hoc Networks (VANETs)," Jun. 2020. doi: 10.1109/HORA49412.2020.9152928.

[4] A. Guerna, S. Bitam, and C. T. Calafate, "Roadside Unit Deployment in Internet of Vehicles Systems: A Survey," *Sensors*, vol. 22, no. 9. Multidisciplinary Digital Publishing Institute, p. 3190, Apr. 21, 2022. doi: 10.3390/s22093190.

[5] L. Codeca, R. Frank, and T. Engel, "Luxembourg SUMO Traffic (LuST) Scenario: 24 hours of mobility for vehicular networking research," in *IEEE Vehicular Networking Conference, VNC*, Jan. 2016, vol. 2016-Janua, pp. 1–8. doi: 10.1109/VNC.2015.7385539.

[6] L. Codeca and J. Harri, "Towards multimodal mobility simulation of C-ITS: The Monaco SUMO traffic scenario," in *IEEE Vehicular Networking Conference, VNC*, Jan. 2018, vol. 2018-Janua, pp. 97–100. doi: 10.1109/VNC.2017.8275627.

[7] S. T. Rakkesh, A. R. Weerasinghe, and R. A. C. Ranasinghe, "A decentralized vehicle re-routing approach using vehicular ad-hoc networks," in *16th International Conference on Advances in ICT for Emerging Regions, ICTer 2016 - Conference Proceedings*, Jan. 2017, pp. 201–207. doi: 10.1109/ICTER.2016.7829919.

[8] S. Talib Hasoon and M. A. Mahdi, "A Developed Realistic Urban Road Traffic in Erbil City Using Bi-directionally Coupled Simulations," *Qalaai Zanist Sci. J.*, vol. 2, no. 2, pp. 27–34, Apr. 2017, doi: 10.25212/lfu.qzj.2.2.03.

[9] A. Varga, "A practical introduction to the OMNeT++ simulation framework," in *EAI/Springer Innovations in Communication and Computing*, Springer Science and Business Media Deutschland GmbH, 2019, pp. 3–51. doi: 10.1007/978-3-030-12842-5_1.

[10] M. J. Haidari and Z. Yetgin, "Veins based studies for vehicular ad hoc networks," Sep. 2019. doi: 10.1109/IDAP.2019.8875954.

[11] Y. Regragui and N. Moussa, "Investigating the impact of real-time path planning on reducing vehicles traveling time," in *Proceedings - 2018 International Conference on Advanced Communication*

*Technologies and Networking, CommNet 2018*, May 2018, pp. 1–6. doi: 10.1109/COMMNET.2018.8360277.

[12] M. Al Shareeda, A. Khalil, and W. Fahs, "Realistic heterogeneous genetic-based RSU placement solution for V2I networks," *Int. Arab J. Inf. Technol.*, vol. 16, no. 3ASpecial Issue, pp. 540–547, 2019.

[13] F. G. Abdulkadhim, Z. Yi, A. N. Onaizah, F. Rabee, and A. M. A. Al-Muqarm, "Optimizing the Roadside Unit Deployment Mechanism in VANET with Efficient Protocol to Prevent Data Loss," *Wirel. Pers. Commun.*, pp. 1–29, May 2021, doi: 10.1007/s11277-021-08410-6.

[14] L. Jiang, T. G. Molnár, and G. Orosz, "On the deployment of V2X roadside units for traffic prediction," *Transp. Res. Part C Emerg. Technol.*, vol. 129, p. 103238, Aug. 2021, doi: 10.1016/j.trc.2021.103238.

[15] J. Um, "Performance Analysis According to RSU Range of VANET-based Communication Vehicle," *Int. J. Sci. Eng. Smart Veh.*, vol. 4, no. 1, pp. 1–6, 2020, doi: 10.21742/ijsesv.2020.4.1.01.

[16] S. Ben Chaabene, T. Yeferny, and S. Ben Yahia, "A Roadside Unit Placement Scheme for Vehicular Ad-hoc Networks," in *Advances in Intelligent Systems and Computing*, 2020, vol. 926, pp. 619–630. doi: 10.1007/978-3-030-15032-7_52.

[17] V. Padmapriya, A. K. Ashok, D. N. Sujatha, and K. R. Venugopal, "Road Side Unit assisted Emergency Vehicle Transit Approach for Urban Roads using VANET," Feb. 2019. doi: 10.1109/ICECCT.2019.8869527.

[18] S. Ma, F. Wen, X. Zhao, Z. M. Wang, and D. Yang, "An Efficient V2X Based Vehicle Localization Using Single RSU and Single Receiver," *IEEE Access*, vol. 7, pp. 46114–46121, 2019, doi: 10.1109/ACCESS.2019.2909796.

[19] J. Wu, M. Fang, H. Li, and X. Li, "RSU-Assisted Traffic-Aware Routing Based on Reinforcement Learning for Urban Vanets," *IEEE Access*, vol. 8, pp. 5733–5748, 2020, doi: 10.1109/ACCESS.2020.2963850.

[20] H. Yang, Z. Jia, and G. Xie, "Delay-bounded and cost-limited RSU deployment in Urban Vehicular Ad Hoc Networks," *Sensors (Switzerland)*, vol. 18, no. 9, p. 2764, Aug. 2018, doi: 10.3390/s18092764.

[21] L. Luo, L. Sheng, H. Yu, and G. Sun, "Intersection-Based V2X Routing via Reinforcement Learning in Vehicular Ad Hoc Networks," *IEEE Trans. Intell. Transp. Syst.*, vol. 23, no. 6, pp. 5446–5459, Jun. 2022, doi: 10.1109/TITS.2021.3053958.

[22] C. Sommer *et al.*, "Veins: The open source vehicular network simulation framework," in *EAI/Springer Innovations in Communication and Computing*, Springer Science and Business Media Deutschland GmbH, 2019, pp. 215–252. doi: 10.1007/978-3-030-12842-5_6.

[23] F. Arena, G. Pau, and A. Severino, "A review on IEEE 802.11p for intelligent transportation systems," *Journal of Sensor and Actuator Networks*, vol. 9, no. 2. Multidisciplinary Digital Publishing Institute, p. 22, Apr. 26, 2020. doi: 10.3390/jsan9020022.